\pdfoutput=1
\PassOptionsToPackage{bookmarks=false,colorlinks=true,linkcolor=blue,urlcolor=cyan,filecolor=green,citecolor=magenta,pdfstartview=FitV,hyperfootnotes=false,pdftitle={Angle of Null Energy Condition Lines in Critical Spacetimes},pdfauthor={Christian Ecker, Florian Ecker, Daniel Grumiller},pdfsubject={Angle of Null Energy Condition Lines in Critical Spacetimes},pdfkeywords={Angle of Null Energy Condition Lines in Critical Spacetimes},bookmarksopen=true}{hyperref}
\documentclass[aps,prl,10pt,twocolumn,superscriptaddress,amsmath,amssymb]{revtex4-2}

\usepackage{graphicx,bm,braket,mathrsfs,orcidlink,pgfplots}
\usepackage[normalem]{ulem}
\usepackage{tikz}
\usetikzlibrary{decorations.pathmorphing,backgrounds,calc,shapes,arrows,shapes.misc,shadows}
\tikzset{cross/.style={cross out, draw=black, minimum size=2*(#1-\pgflinewidth), inner sep=0pt, outer sep=0pt}, cross/.default={1pt}}
\usepgfplotslibrary{fillbetween}

\DeclareMathOperator{\extdm}{d}
\newcommand{\extd}{\extdm \!}
\newcommand{\arccot}{\textrm{arccot}}

\begin{document}


\preprint{TUW-24-07}

\def\mytitle{Angle of Null Energy Condition Lines in Critical Spacetimes}
\title{\mytitle}

\author{\orcidlink{0000-0002-8669-4300}Christian Ecker}
\email{ecker@itp.uni-frankfurt.de}
\affiliation{%
Institute for Theoretical Physics, Goethe University\\ 60438, Frankfurt am Main, Germany
}%

\author{\orcidlink{0000-0002-0449-0081}Florian Ecker}
\email{fecker@hep.itp.tuwien.ac.at}
\affiliation{%
Institute for Theoretical Physics, TU Wien\\
Wiedner Hauptstrasse 8–10/136, A-1040 Vienna, Austria
}

\author{\orcidlink{0000-0001-7980-5394}Daniel Grumiller}
\email{grumil@hep.itp.tuwien.ac.at}
\affiliation{%
Institute for Theoretical Physics, TU Wien\\
Wiedner Hauptstrasse 8–10/136, A-1040 Vienna, Austria
}

\begin{abstract}
We identify a new critical parameter in Choptuik’s gravitational collapse: the angle at which null energy condition (NEC) saturation lines intersect at the center of the critical spacetime. These NEC lines coincide with regions of vanishing curvature, dividing spacetime into stripes of positive and negative curvature. By numerically solving Choptuik’s original system we find the NEC angle to be $\alpha\approx0.64$ ($\approx 37^\circ$) and analytically derive $\alpha=2\,\arccot(D-1)$ for any spacetime dimension $D>3$.
\end{abstract}

\maketitle


\section{Introduction and main result}
\label{sec:1}

Black holes have paradoxical properties --- once they have settled down to a stationary state, they are the simplest macroscopic objects that one could imagine since, classically, all their properties are determined by a handful of numbers, including their mass and angular momentum (see, e.g., \cite{Chrusciel:2020fql}). Non-stationary situations like the collapse or evaporation of black holes are vastly more complicated  (see \cite{Grumiller:2022qhx} and Refs.~therein). In this work, we focus on the spherically symmetric collapse of matter to a black hole. 

Critical collapse, discovered numerically by Choptuik \cite{Choptuik:1993jv} for four-dimensional (4D) spherically symmetric general relativity minimally coupled to a massless scalar field \cite{Christodoulou:1986zr}, is characterized by two critical parameters (further explicated below): the universal exponent $\gamma$ that governs various power-law scalings, and the echoing period $\Delta$ of the critical solution at the threshold of black hole formation. In 4D, Choptuik found $\gamma\approx0.37$ and $\Delta\approx3.44$. There is no known way to derive these two numbers from first principles without numerically solving a system of coupled non-linear partial differential equations (PDEs).

Our Letter aims to provide a third universal number that characterizes the critical solution, the NEC angle $\alpha\approx 0.64$, to be defined below, and to calculate it from first principles in arbitrary spacetime dimensions $D>3$. Our main result is 
\begin{equation}
\alpha=2\,\arccot(D-1)\,.
\label{eq:NEC1}
\end{equation}
 
\section{Critical collapse}
\label{sec:2}

Originally, Choptuik arrived at his seminal results by considering a variety of 1-parameter families of initial data, labeled by a family parameter $p\in[0,1]$, with the convention that for $p$ close to zero, no black hole forms and for $p$ close to one always a black hole forms. For each such family, there is a critical value $p=p_\ast$ that describes a critical spacetime solution at the threshold of black hole formation --- it is neither a black hole nor a regular spacetime but rather a naked singularity with three remarkable properties (see \cite{Gundlach:2002sx} and Refs.~therein):
\begin{enumerate}
\item It is universal, i.e., always the same spacetime, regardless of the specific family. Constructing the critical solution is akin to solving critical collapse.
\item It is almost stable under linearized perturbations but has one unstable mode, thereby allowing a renormalization group flow-inspired interpretation as an attractor of codimension-1 \cite{Koike:1995jm}. The reciprocal of the eigenvalue of the unstable mode is the Choptuik parameter $\gamma$ \cite{Gundlach:1995kd}. The critical solution is analogous to a prethermalized state \cite{Berges:2004ce,Langen:2016vdb,Berges:2020fwq}.
\item It features discrete self-similarity (DSS). In adapted coordinates, this means the metric can be written as $g_{\mu\nu}(\tau,x^i)=e^{-2\tau}\tilde g_{\mu\nu}(\tau,x^i)$ with $\tilde g_{\mu\nu}(\tau+\Delta,x^i)=\tilde g_{\mu\nu}(\tau,x^i)$, where $\Delta$ is the echoing period. The ``crystal coordinate'' $\tau$ defines a crystal vector $c^\mu\partial_\mu=\partial_\tau$ that can be time-, light-, or spacelike, depending on the region of the critical solution. The metric $\tilde g$, therefore, describes a spacetime crystal --- a configuration that spontaneously breaks spacetime translations to a discrete subgroup --- and the critical spacetime is conformally related to this spacetime crystal. 
\end{enumerate}

While the critical solution fully captures the information about Choptuik's critical parameters, they also feature in near-critical solutions, which is how they were discovered at first. For example, above the threshold of black hole formation ($p>p_\ast$), the final mass $M$ of the black hole produced by the spherically symmetric collapse of a scalar field behaves like
\begin{equation}
M \propto (p-p_\ast)^\gamma
    \label{eq:NEC2}
\end{equation}
where the proportionality constant and the value $p_\ast$ are family-dependent, but the exponent $\gamma$ is universal. 

An efficient way to determine the critical parameters is to assume there is a critical solution with DSS and then construct it numerically. As explained in \cite{Gundlach:1995kd}, the spherically symmetric Einstein--Klein--Gordon-system turns into an eigenvalue problem, the solution of which yields the echoing period $\Delta$. Moreover, linearized perturbations around this critical solution lead to another eigenvalue problem, the solution of which yields the scaling exponent $\gamma$, see the discussion in \cite{Gundlach:1996eg}.

\section{Critical solution in dimension \texorpdfstring{$\boldsymbol{D>3}$}{D>3}}
\label{sec:3}

Our first step is to generalize the construction of the critical solution to arbitrary dimensions $D>3$ such that $D$ enters as an arbitrary parameter of the field equations, allowing analytical continuation to non-integer values. This differs from the higher-dimensional approaches in \cite{Sorkin:2005vz,Bland:2005kk} who followed Choptuik's numerical construction.

The field equations descending from the Einstein--Klein--Gordon action in $D>3$ spacetime dimensions (with Newton's constant $G^{(D)}$),
\begin{multline}
    S\big[g^{(D)}_{\mu\nu},\,\psi\big]=\frac{1}{16\pi G^{(D)}}\int \extd^Dx\sqrt{-g^{(D)}}\, R^{(D)}\\
    -\frac{1}{2}\int \extd^Dx\sqrt{-g^{(D)}}\, g_{(D)}^{\mu \nu}(\partial_\mu \psi)(\partial_\nu \psi) 
\end{multline}
simplify to a two-dimensional (2D) system of PDEs [that we display below in Eqs.~\eqref{eq:eom1}-\eqref{eq:eom5}] when imposing spherical symmetry and DSS.
 
In coordinates adapted to these continuous and discrete symmetries, the $D$-dimensional metric
\begin{equation}
     \extd s^2_{(D)}=e^{-2\tau}\Big(\tilde g_{\alpha\beta}(\tau,x)\,\extd x^\alpha\extd x^\beta+x^2\extd \Omega^2_{S^{D-2}}\Big)
     \label{eq:whynolabel}
\end{equation}
is essentially governed by the 2D spacetime crystal metric $\tilde g_{\alpha\beta}(\tau+\Delta,x)=\tilde g_{\alpha\beta}(\tau,x)$, which depends on the crystal coordinate $\tau$ and another coordinate denoted by $x$. The line element of the round $(D-2)$-sphere, $\extd \Omega^2_{S^{D-2}}$, plays no active role in our analysis. The locus $x\to 0$ corresponds to the center of spacetime since the size of the $(D-2)$-sphere shrinks to zero.

\newcommand{\cf}{\omega}

We follow the gauge choices of \cite{Martin-Garcia:2003xgm} so that the spacetime crystal metric
\begin{equation}
\tilde g_{\alpha\beta}\extd x^\alpha\extd x^\beta = e^{\cf}\,\big((x^2-f^2)\extd\tau^2-2x\extd\tau \extd x+\extd x^2\big)
\label{eq:lalapetz}
\end{equation}
depends on two periodic functions, $\cf(\tau+\Delta,x)=\cf(\tau,x)$ and $f(\tau+\Delta,x)=f(\tau,x)$, with an echoing period $\Delta$ that depends on the dimension $D$. In the relevant patch (see Fig.~\ref{fig:1} below), the functions obey the inequalities $\cf\geq 0$ and $0<f\leq 1$.

Using $x$ as the evolution parameter, the Einstein--Klein--Gordon field equations with the choices above reduce to four evolution equations
\begin{align}
    x\,\partial_x \cf &= (D-3)\big(1-e^{\cf}\big)+\frac{1}{2}(U^2+V^2)\label{eq:eom1}\\
    x\,\partial_x f &= (D-3)\big(e^{\cf}-1\big)f\label{eq:eom2}\\
    \frac{2x}{f}\,v^\mu\partial_\mu U &= \big(D-2+2(3-D)e^{\cf}\big)U+(D-2)V\label{eq:eom3}\\
    \frac{2x}{f}\,u^\mu\partial_\mu V  &= \big(2-D+2(D-3)e^{\cf}\big)V+(2-D)U\label{eq:eom4}
\end{align}
and one constraint equation
\begin{equation}
\partial_\tau\cf=\frac{(f-x)V^2-(f+x)U^2}{2x}+(D-3)\big(e^{\cf}-1\big) \label{eq:eom5}
\end{equation}
where we used the first-order matter variables
\begin{equation}
V=\frac{\sqrt{8\pi G^{(D)}}x}{\sqrt{D-2}\,f}v^\mu\partial_\mu \psi\qquad U=\frac{\sqrt{8\pi G^{(D)}}x}{\sqrt{D-2}\,f}u^\mu\partial_\mu \psi
\label{eq:NEC3}
\end{equation}
defined by the null vector fields
\begin{equation}
     v^\mu\partial_\mu=\partial_\tau+(f+x)\partial_x \qquad\quad u^\mu\partial_\mu=\partial_\tau-(f-x)\partial_x\,. 
\end{equation}

We now discuss boundary conditions. At the center of the spacetime, $x=0$, we demand that the $D$-dimensional Ricci-scalar,
\begin{multline}
 R^{(D)} \big|_{x\ll 1}=\mathcal{O}(x^0)+\big(1-e^{-\cf}\big)_{x=0}\frac{(D-2)(D-3)e^{2\tau}}{x^2}\\
 +\big[e^{-\cf}\big((D-3)\partial_x\cf-2\partial_x\ln{f}\big)\big]_{x=0} \frac{(D-2)e^{2\tau}}{x} 
    \label{eq:RD}
\end{multline}
remains finite for any finite value of the crystal time $\tau$, yielding the regularity condition $\cf(\tau,0)=0$ that eliminates the second-order pole in $x$. Taylor-expanding the equations of motion \eqref{eq:eom1}-\eqref{eq:eom4} to second order near the origin $x=0$ implies additionally $\partial_x\cf(\tau,0)=\partial_xf(\tau,0)=0$ and $U(\tau,0)=V(\tau,0)=0$, $\partial_xU(\tau,0)=\partial_xV(\tau,0)$, $\partial_x^2U(\tau,0)=-\partial_x^2V(\tau,0)$, 
ensuring that also the first order pole in $x$ in the Ricci scalar \eqref{eq:RD} is absent on-shell.

The other evolution boundary, $x=1$, is called ``self-similar horizon'' (SSH) and, by choice, corresponds to the line where the crystal vector $c^\mu\partial_\mu=\partial_\tau$ becomes null, $c^2=0$, which in the gauge \eqref{eq:lalapetz} is captured by the condition $g_{\tau\tau}|_{x=1}=0$. Therefore, we need to impose the SSH boundary condition $f(\tau,1)=1$. The patch of the spacetime crystal described above is depicted as the shaded region in Fig.~\ref{fig:1}. It corresponds to the fundamental domain of the ``past patch'' in the nomenclature of \cite{Martin-Garcia:2003xgm}. 

\begin{figure}[htb]
\centering
\begin{tikzpicture}[xscale=3.0,yscale=3.0]
\draw (0,0) coordinate (ori) -- (0,2) coordinate (sin) node[left, yshift=1mm] at (ori) {$\tau\to-\infty$};
\draw (ori) -- (2,2) coordinate (i0); 
\draw (i0) -- (1,3) coordinate (iplus) node[left] at (i0) {$\dots$};
\draw[thin,decorate,decoration={zigzag, amplitude=1pt, segment length=3pt}] (iplus) -- (sin) node[midway, above, sloped] {Cauchy horizon};
\fill (sin) circle (1pt);
\draw[thin,dashed] (sin) -- (1,1) coordinate (ssh) node[left] at (sin) {$\tau\to\infty$} node[right] at (ssh) {SSH} node[pos=0.62, above, sloped] {$\scriptstyle x=1$};
\draw[thin,dotted] (0,0.03125) coordinate (lo5) -- (0.983375,1.015625) coordinate (tm5);
\draw[thin,draw=none] (0,0.0625) coordinate (lo4x) -- (0.96875,1.03125) coordinate (tm4);
\draw[thin,dotted] (0,0.05625) coordinate (lo4) -- (0.971875,1.028125) coordinate (tm4x);
\draw[thin,draw=none] (0,0.125) coordinate (lo3x) -- (0.9375,1.0625) coordinate (tm3);
\draw[thin,dotted] (0,0.1125) coordinate (lo3) -- (0.94375,1.05625) coordinate (tm3x) node[left] at (lo3) {$\dots$};
\draw[thin,draw=none] (0,0.25) coordinate (lo2x) -- (0.875,1.125) coordinate (tm2);
\draw[thin,dotted] (0,0.225) coordinate (lo2) -- (0.8875,1.1125) coordinate (tm2x) node[left] at (lo2) {$\tau=-2\Delta$};
\draw[thin,draw=none] (0,0.5) coordinate (lo1x) -- (0.75,1.25) coordinate (tm1);
\draw[thin,dotted] (0,0.45) coordinate (lo1) -- (0.775,1.225) coordinate (tm1x) node[left] at (lo1) {$\tau=-\Delta$};
\draw[thin,draw=none] (0,1) coordinate (lox) -- (0.5,1.5) coordinate (t0);
\draw[thin,dotted] (0,0.9) coordinate (lo) -- (0.55,1.45) coordinate (t0x) node[left] at (lo) {$\tau=0$};
\draw[thin,draw=none] (0,1.5) coordinate (hix) -- (0.25,1.75) coordinate (t1);
\draw[thin,dotted] (0,1.4) coordinate (hi) -- (0.3,1.7) coordinate (t1x) node[left] at (hi) {$\tau=\Delta$};
\draw[thin,draw=none] (0,1.75) coordinate (hi2x) -- (0.125,1.875) coordinate (t2);
\draw[thin,dotted] (0,1.7) coordinate (hi2) -- (0.15,1.85) coordinate (t2x) node[left] at (hi2) {$\tau=2\Delta$};
\draw[thin,draw=none] (0,1.875) coordinate (hi3x) -- (0.0625,1.9375) coordinate (t3);
\draw[thin,dotted] (0,1.85) coordinate (hi3) -- (0.075,1.925) coordinate (t3x) node[left] at (hi3) {$\dots$};
\draw[thin,draw=none] (0,1.9375) coordinate (hi4x) -- (0.03125,1.96875) coordinate (t4);
\draw[thin,dotted] (0,1.925) coordinate (hi4) -- (0.0325,1.9675) coordinate (t4);
\draw[thin,dotted] (0,1.96875) coordinate (hi5) -- (0.015625,1.984375) coordinate (t5);
\draw[thin] (lo) -- (tm1) node[below, midway, sloped] {$\scriptstyle\tau=0$};
\draw[thin] (hi) -- (t0)  node[above, midway, sloped] {$\scriptstyle\tau=\Delta$};
\fill[gray, opacity=0.6] (lo) -- (tm1) -- (t0) -- (hi) -- cycle;
\draw[thin] (hi5) -- (t4);
\draw[thin] (hi4) -- (t3);
\draw[thin] (hi3) -- (t2);
\draw[thin] (hi2) -- (t1);
\draw[thin] (lo1) -- (tm2);
\draw[thin] (lo2) -- (tm3);
\draw[thin] (lo3) -- (tm4);
\draw[thin] (lo4) -- (tm5);
\draw[thin] (t4) .. controls (0.0625,2.0) .. (0.03125,2.03125);
\draw[thin] (t3) .. controls (0.125,2.0) .. (0.0625,2.0625);
\draw[thin] (t2) .. controls (0.25,2.0) .. (0.125,2.125);
\draw[thin] (t1) .. controls (0.5,2.0) .. (0.25,2.25);
\draw[thin] (t0) .. controls (1.0,2.0) .. (0.5,2.5) coordinate (end0) node[above, midway, sloped] {$\scriptstyle \tau=\Delta$};
\draw[thin] (tm1) .. controls (1.5,2.0) .. (0.75,2.75) coordinate (endm1)  node[above, midway, sloped] {$\scriptstyle \tau=0$};
\draw[thin] (tm2) .. controls (1.75,2.0) .. (0.875,2.875);
\draw[thin] (tm3) .. controls (1.875,2.0) .. (0.9375,2.9375);
\draw[thin] (tm4) .. controls (1.9375,2.0) .. (0.96875,2.96875);
\draw[thin] (tm5) .. controls (1.96875,2.0) .. (0.984375,2.984375);
\begin{scope}
    \fill[gray, opacity=0.1] 
        (tm1) .. controls (1.5,2.0) .. (endm1) -- 
        (end0) .. controls (1.0,2.0) .. (t0) -- cycle;
\end{scope}
\draw[thin] (lo) -- (hi) node[above, midway, sloped] {$\scriptstyle x=0$};
\node at (0,2.6) {singularity};
\draw[->, thin] (0,2.55) -- (0,2.05);
\draw[domain=0:0.95, smooth, variable=\x, thin, black, dotted] plot ({0.9+\x}, {0.1+0.3*\x/sqrt((\x)*(\x)+0.25)});
\fill[gray, opacity=0.6] (0.9,0.1) coordinate (pA) -- (1.85,0.1) coordinate (pB) -- (1.85,0.7) coordinate (pC) -- (0.9,0.7) coordinate (pD) -- cycle;
\draw[thin] (pA) -- (pB) node[below, midway] {$\scriptstyle \tau=0$};
\draw[thin] (pB) -- (pC) node[left, pos=0.6] {$\scriptstyle f=1$} node[left, pos=0.4] {$\scriptstyle \mathcal{D}_\tau V=U$} node[above, midway, sloped, rotate=180] {$\scriptstyle x=1$};
\draw[thin] (pC) -- (pD) node[above, midway] {$\scriptstyle \tau=\Delta$};
\draw[thin] (pD) -- (pA) node[right, pos=0.2] {$\scriptstyle \cf=U=V=0$} node[right, pos=0.4] {$\scriptstyle \partial_x\cf=\partial_xf=0$} node[right, pos=0.6] {$\scriptstyle \partial_xU=\partial_xV$}  node[right, pos=0.8] {$\scriptstyle \partial_x^2U=-\partial^2_xV$} node[above, midway, sloped, rotate=180] {$\scriptstyle x=0$};
\draw[<->, thin] (1.7, 0.09) .. controls (2.1, -0.22) and (2.1, 1.02) .. (1.7, 0.71) node[above, pos=0.85] {\small identify};
\draw[->, thin, gray, opacity=0.5] (0.05,0.85) -- (0.85,0.1);
\end{tikzpicture}
\caption{Penrose diagram of critical spacetime with DSS. The past patch lies between the origin ($x=0$) and the SSH ($x=1$, dashed line) where $f=1$ in the gauge \eqref{eq:lalapetz}. The crystal metric $\tilde g$ is identified periodically along the displayed $\tau=\rm const.$ lines. A fundamental domain is highlighted as shaded region, spanning the area in the intervals $x\in[0,1]$ and $\tau\in[0,\Delta]$. Dotted lines are null curves. The outer patch (of no relevance in our analysis) is the diamond-shaped region bounded by the SSH, the Cauchy horizon emanating from the singularity at $x=0,\tau\to\infty$, and future/past null infinity. The crystal coordinate $\tau$ is timelike in the past patch, lightlike on the SSH, and spacelike in the outer patch. The inset on the bottom right shows the fundamental domain in the past patch with boundary conditions at origin and SSH.}
\label{fig:1}
\end{figure}
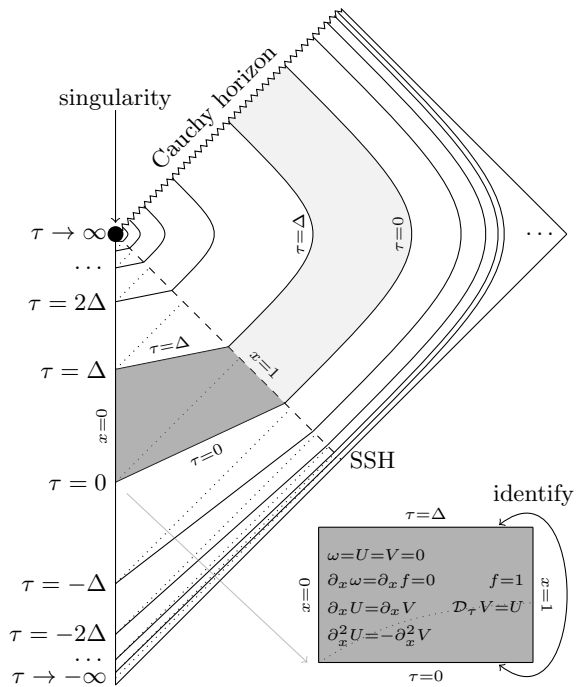

A final SSH boundary condition stems from the regularity of the PDE \eqref{eq:eom4} at $x=f=1$, namely $\mathcal{D}_{\tau}V(\tau,1)=U(\tau,1)$ with $\mathcal{D}_\tau:=\frac{2}{2-D}\partial_\tau+2\frac{D-3}{D-2}e^{\cf(\tau,1)}-1$. 

Solving the PDE system \eqref{eq:eom1}-\eqref{eq:eom5} with the boundary conditions above can be done numerically at any finite $D>3$ and leads to unique solutions for the four functions $\cf,f,U,V$ and the echoing period $\Delta$. How to do so will be described elsewhere in detail \cite{prep}, together with the results for $\Delta$ and $\gamma$, and closely follows the algorithm explained in \cite{Martin-Garcia:2003xgm}. 

For our purposes, we only need some aspects of the critical solution near the center. To explain why this is so, we focus first on the well-known numerical results for $D=4$ and display the selection of them that is relevant for our analysis. More importantly, we shall identify a new gauge invariant observable of the critical spacetime, the NEC angle.

\section{Numerical results for NEC angle in 4D}
\label{sec:4}

We recovered the critical solution constructed in \cite{Martin-Garcia:2003xgm} by using essentially their numerical algorithm. Numerical details and the generalization to arbitrary dimensions will be discussed elsewhere \cite{prep}. Here, we only need a few details from this analysis.

\begin{figure}[htb]
\centering
\includegraphics[width=0.48\textwidth]{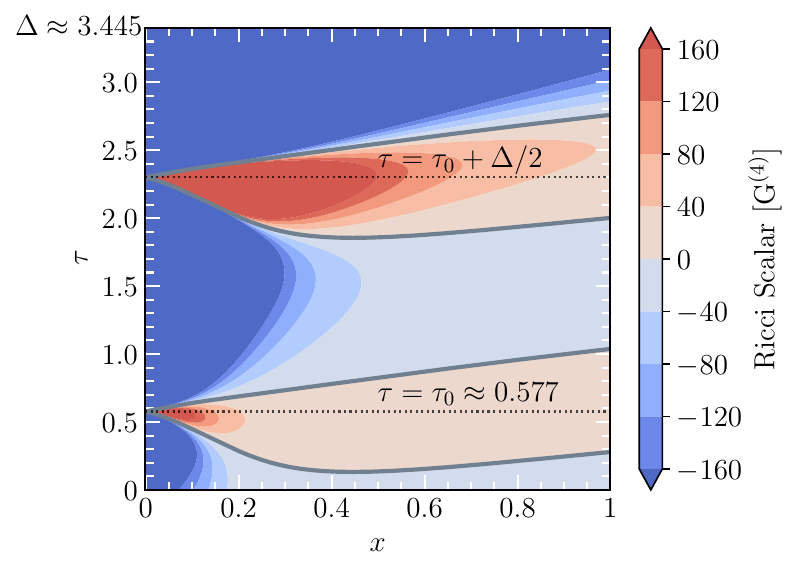}
\caption{Ricci scalar in fundamental domain. 
Zero curvature contours marked by gray lines. Values $\tau=\tau_0\approx0.577$, $\tau=\tau_0+\tfrac\Delta2$ marked by dotted lines bisect vertex angles at $x=0$.}
\label{fig:2}
\end{figure}

Since we are interested in diffeomorphism invariant observables, in Fig.~\ref{fig:2} we display the Ricci scalar $R^{(D)}$ in the fundamental domain $\tau\in[0,\Delta]$ in the past patch $x\in[0,1]$.
Remarkably, the critical spacetime has stripes of positive and negative curvature. There are two such stripes in each fundamental domain because the Ricci scalar is quadratic in the scalar field, see Eq.~\eqref{eq:R} below. The lines of vanishing curvature separating the stripes emanate from the origin, where they coalesce into a vertex with a certain opening angle. 

With respect to the proper length and proper time near the vertex, this opening angle is a diffeomorphism invariant number and hence constitutes another critical parameter that characterizes this critical spacetime. Alternatively, this opening angle can be understood in terms of the relative boost between the two spacelike hypersurfaces of vanishing curvature. We shall use this alternative in our analytical derivation in the next Section. 

Before evaluating this new critical parameter, we briefly switch gears and consider the classical NEC. The NEC inequalities
$T_{\mu\nu} v^\mu v^\nu= (v^\mu\partial_\mu\psi)^2  \geq 0$, $T_{\mu\nu} u^\mu u^\nu = (u^\mu\partial_\mu\psi)^2 \geq 0$
trivially hold everywhere. A less trivial aspect of our solutions is that one of the NEC inequalities saturates when $U/x$ or $V/x$ defined in \eqref{eq:NEC3} vanishes.
\begin{equation}
\textrm{NEC-saturation:}\qquad \frac{UV}{x^2} = 0
    \label{eq:NEC5}
\end{equation}

The NEC saturation lines are identical to the gray lines in Fig.~\ref{fig:2} since the Ricci scalar can be expressed as
\begin{equation}
R^{(D)} = -(D-2)\,e^{2\tau}e^{-\cf}\,\frac{UV}{x^2}\,.
\label{eq:R}
\end{equation}
Therefore, the new critical parameter uncovered above has an alternative interpretation as the opening angle of the two NEC saturation lines in Fig.~\ref{fig:2} at the origin $x=0$. We refer to this critical parameter as ``NEC angle'' and denote it by $\alpha$.

\begin{figure}[htb]
\centering
\includegraphics[width=0.5\textwidth]{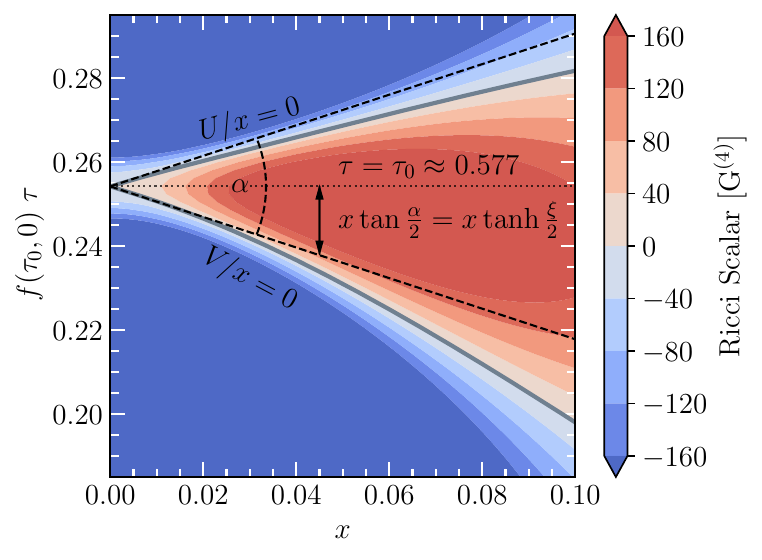}
\caption{NEC angle $\alpha\approx37^\circ$ (dashed lines) in 4D critical spacetime. To get a diffeomorphism invariant angle, we rescaled to proper time $t=f(\tau_0,0)\,\tau$ with $f(\tau_0,0)\approx0.441$ [crystal metric \eqref{eq:lalapetz} is conformal to ${\extd}\,s^2=-{\extd}\,t^2+{\extd}\,x^2+\mathcal{O}(x)+\mathcal{O}(\tau-\tau_0)$]. Dotted: $\tau=\tau_0\approx0.577$. Displayed in gray: NEC lines $U/x=0$ and $V/x=0$. Formula: Lorentzian relation between NEC angle $\alpha$ and relative rapidity $\xi$ [see Eq.~\eqref{eq:angelinajolie}].}
\label{fig:4}
\end{figure}

To better resolve the precise value of the NEC angle, we zoom into the $x=0$ region in Fig.~\ref{fig:4} near $\tau=\tau_0$. Measuring the NEC angle we roughly get
\begin{equation}
    \alpha \approx 37^\circ \simeq 0.64\,.
    \label{eq:NEC4}
\end{equation}

In the remainder of our work, we derive this NEC angle analytically for arbitrary spacetime dimensions $D>3$.

\section{Derivation of NEC angle}
\label{sec:6}

The NEC angle is localized at the origin. To derive it, it is sufficient to Taylor-expand the solution for the critical spacetime around the origin to quadratic order in $x$, where we assume that the critical solution has appropriate analyticity properties to allow such an expansion~\footnote{%
Pragmatically, we find this to be true by numerically constructing the critical solution and comparing it with a solution obtained by Taylor expansion, which works well up to the expected numerical accuracy \cite{prep}. However, we have no proof of Taylor-expandability around $x=0$ for arbitrary $D>3$.}. 
Consistency with the boundary conditions and equations of motion leads to 
\begin{align}
    \cf(\tau,x) &= \cf_2(\tau) x^2 + {\cal O}(x^4)\\
    f(\tau,x) &= f_0(\tau) + f_2(\tau) x^2 + {\cal O}(x^4)\\
    U(\tau,x) &= v_1(\tau) x - v_2(\tau) x^2 + {\cal O}(x^3)\\
    V(\tau,x) &= v_1(\tau) x + v_2(\tau) x^2 + {\cal O}(x^3)\,.
\end{align}
The first two coefficient functions of the matter field variables $U$ and $V$ are identical to each other up to signs.

Inserting the expansions above into the equations of motion \eqref{eq:eom1}-\eqref{eq:eom5} yields five (partly redundant) conditions for the three functions $\cf_2(\tau),f_2(\tau),v_2(\tau)$ solved by
\begin{equation}
\cf_2=\frac{v_1^2}{D-1}\quad f_2=\frac{(D-3)f_0v_1^2}{2(D-1)}\quad v_2=\frac{v_1+\dot v_1}{(D-1)f_0}
    \label{eq:NEC7}
\end{equation}
where dot denotes derivative with respect to $\tau$.

Since the boundary function $v_1(\tau)$ features prominently in the solution \eqref{eq:NEC7}, we discuss its key property: It has to be a (non-trivial) periodic function with zero average. Periodicity follows from our construction, while the zero average property follows from the definition \eqref{eq:NEC3} provided the scalar field $\psi$ is periodic (rather than quasi-periodic). While generalizations are conceivable where the scalar field is only quasi-periodic and has some winding mode in $\tau$, the results in $D=4$ show that no such winding mode is switched on in the critical solution (see, e.g., \cite{Gundlach:2002sx}). We assume the same to be true for any $D>3$.

The boundary data $v_1(\tau)$ are therefore given by some sinusoidal curve. The precise shape of that curve is irrelevant for our purposes. The only aspect that matters is that any non-trivial sinusoidal curve with zero average must have at least two zeros (of odd degree) in the fundamental domain \footnote{%
Numerical simulations in $D>3$ show that these sinusoidal curves have exactly two zeros of degree one in the fundamental domain \cite{prep}.}. Generically, these zeros are simple, i.e., of first degree.

This observation is crucial since the two NEC lines $U/x=0=V/x$ at the origin coalesce to a NEC vertex at $x=0$ given by the condition $v_1(\tau_0) = 0$.

To get the NEC angle, we linearize in $\tau$ around such a zero at $\tau=\tau_0$, $v_1(\tau) \approx v_0 (\tau-\tau_0) + {\cal O}(\tau-\tau_0)^2$ with $v_0\neq 0$, and solve the equations for the two NEC lines $U/x=0$ and $V/x=0$ to linear order in $x$ and $\tau-\tau_0$. This leads to two 2D vectors,
\begin{equation}
n_\pm^\alpha=\begin{pmatrix} 
\pm 1\\ f_0(\tau_0)(D-1) \end{pmatrix}    
\end{equation}
that point in the direction of the NEC lines away from the origin. For $D>2$, both tangent vectors $n_\pm$ are spacelike.

The (normalized) inner product between these vectors (evaluated at the NEC vertex $\tau=\tau_0$, $x=0$) gives the relative rapidity $\xi$ between the two NEC lines,
\begin{equation}
    \cosh\xi=\frac{n_+^\alpha n_-^\beta g_{\alpha\beta}}{\sqrt{n_+^\gamma n_+^\delta g_{\gamma\delta}\,n_-^\alpha n_-^\beta g_{\alpha\beta}}}=\frac{D^2-2D+2}{D(D-2)} \,.
    \label{eq:angelinajolie}
\end{equation}
Remarkably, the dependence on the initial data $f_0$ and $v_0$ drops out in the result \eqref{eq:angelinajolie}, which only depends on the spacetime dimension $D$. Solving \eqref{eq:angelinajolie} for the rapidity yields the elementary expression
\begin{equation}
\xi = \ln\frac{D}{D-2}\,.
    \label{eq:NEC9}
\end{equation}
Converting rapidity \eqref{eq:NEC9} into a dimensionless relative velocity $\beta=\tanh\xi$ between the two dashed spacelike hypersurfaces in Fig.~\ref{fig:4} yields
\begin{equation}
\beta = \frac{2(D-1)}{(D-1)^2+1}\,.
\end{equation} 
The rapidity for $D=4$ is given by $\xi=\ln{2}$ and the velocity by $\beta=0.6$.

The geometric angle $\alpha$ between the two NEC lines is the Gudermannian of the rapidity (see Fig.~\ref{fig:4})~\footnote{%
Interestingly, \eqref{eq:NEC10} coincides with the trivial NEC angle for the massless Klein--Gordon equation on Minkowski space. However, unlike in critical spacetimes, a NEC vertex is not guaranteed, and when it does exist, it lacks a geometric interpretation in terms of zero-curvature contours. In the final Section, we mention generalizations where different critical NEC angles emerge. We thank Carsten Gundlach for discussions on these issues.},
\begin{equation}
\alpha = \textrm{gd}(\xi) = 2\arctan\tanh\frac{\xi}{2} = 2\,\arccot(D-1)\,.
    \label{eq:NEC10}
\end{equation}
Thus, we recover the result announced in \eqref{eq:NEC1}. Evaluating it for $D=4$ yields $\alpha\approx0.64\simeq37^\circ$, compatible with our numerical results from the previous Section.

\section{Outlook}
\label{sec:7}

Since Choptuik’s seminal paper \cite{Choptuik:1993jv}, critical collapse has been studied extensively, primarily with numerical simulations. Most of these studies focused on two quantities, the critical exponent $\gamma$ and the echoing period $\Delta$, assuming various degrees of symmetry, different matter content, and spacetime dimensions.
In this Letter, we revealed that even Choptuik’s original setup of a spherically symmetric massless scalar field still exhibits some previously unexplored features. These include a striped (conformally crystalline) structure in the spacetime curvature and the diffeomorphism invariant opening angle of NEC saturation lines at the center of the critical spacetime, which is analytically accessible in $D>3$.
 
The NEC angle reported in \eqref{eq:NEC10} is a property of critical spacetimes in $D$ spacetime dimensions. However, spacetimes close to the threshold of black hole formation could have properties reflected by this angle, analogous to the dependence of mass just above the threshold \eqref{eq:NEC2}. 

In this Letter, we worked with general relativity as gravitational theory. However, the equations of motion \eqref{eq:eom1}-\eqref{eq:eom5} also apply to 2D dilaton gravity models with scalar matter (see \cite{Grumiller:2002nm,Mertens:2022irh} for reviews and  \cite{Strominger:1993tt,Ashtekar:2010hx} for aspects of critical collapse therein). Thus, our NEC angle discussion also applies to 2D dilaton gravity.

In the large-$D$ limit the NEC angle \eqref{eq:NEC10} simplifies to $\alpha = 2/(D-1) + \mathcal{O}(1/D^3)$. For $D=4$, the large $D$ approximation $\alpha\approx2/3\approx0.67$ gives better than $5\%$ accuracy. It could be illuminating to analyze the system of equations \eqref{eq:eom1}-\eqref{eq:eom5} in this limit, using $1/D$ as small parameter, see e.g.~\cite{Emparan:2013moa,Emparan:2013xia,Rozali:2018yrv,Emparan:2020inr} for some discussions on the large-$D$ limit of general relativity. Conversely, it may be instructive to consider the limit from above $D\to3$, using $D-3$ as a small (positive) expansion parameter.

A straightforward generalization of our results is to keep general relativity with spherical symmetry and DSS but allow for an arbitrary energy-momentum tensor compatible with these symmetries. Regularity and Taylor-expandability at the origin allow to determine the NEC-saturation lines, given by $R^{(D)}_{\mu\nu}v^{\mu}v^\nu=0$ and $R^{(D)}_{\mu\nu}u^{\mu}u^\nu=0$, which coalesce to vertices at the origin. It turns out that, assuming the NEC inequality holds, there will be two NEC-saturation lines that in general are independent of zero-curvature lines, so there is a zoo of possibilities for NEC-angles different from \eqref{eq:NEC10} worthwhile investigating.

Other critical collapse scenarios include generalizations to systems with less symmetry than Choptuik's system \cite{Abrahams:1993wa,Choptuik:2004ha}, different matter content \cite{Evans:1994pj,Liebling:1996dx,Husa:2000kr} or no matter \cite{Ledvinka:2021rve,Baumgarte:2023tdh}, see \cite{Gundlach:2002sx} for a survey and \cite{Baumgarte:2019fai,Marouda:2024epb} (as well as Refs.~therein) for more recent generalizations. It could be rewarding to determine the NEC angle(s) of the corresponding critical solutions and to check under which conditions our formulae \eqref{eq:NEC9}-\eqref{eq:NEC10} apply. 



\paragraph{}\bigskip

\begin{acknowledgments}
We are grateful to Carsten Gundlach for collaboration on numerical constructions of critical solutions in arbitrary spacetime dimensions \cite{prep}. We acknowledge discussions with Roberto Emparan on the large $D$ limit of General Relativity. Additionally, we thank Peter Aichelburg and Piotr Chru\'sciel for discussions.

CE acknowledges support by the DFG through the CRC-TR 211 ``Strong-interaction matter under extreme conditions'' -- project number 315477589 -- TRR 211.
FE and DG were supported by the Austrian Science Fund (FWF), projects P~33789 and P~36619. 
\end{acknowledgments}


\providecommand{\href}[2]{#2}\begingroup\raggedright\endgroup

\end{document}